# Non-monotonic thickness dependence of Curie temperature and ferroelectricity in Two-dimensional SnTe film

Chao Yang, Yanyu Liu, Gang Tang, Xueyun Wang, Jiawang Hong†

*School of Aerospace Engineering, Beijing Institute of Technology, Beijing 100081,*

**Abstract**

Recently, the observation of atomic thin film SnTe with a Curie temperature ($T_c$) higher than that of the bulk (Chang *et. al.*, Science **353**, 274 (2016)) has boosted the research on two-dimensional (2D) ferroic materials tremendously. However, the origin of such phenomenon has yet been thoroughly investigated, which hinder the understanding and design of novel materials with ferroic orders at 2D limit. By using the density functional theory, we investigated the structural and ferroelectrical properties of 2D SnTe, to reveal the thickness dependence. The calculated results demonstrate that the 2D SnTe automatically transform into periodical bilayer structure, resulting from the surface effect. Moreover, based on the double-well potential and atomic distortion analysis, we found the $T_c$ of the 2D SnTe is higher than the bulk counterpart, and more surprisingly, the $T_c$ exhibits an unusual non-monotonous dependence of thickness, featuring a pronounced atomic distortion and Curie temperature maximum at 8 atomic-layers. In addition, this non-monotonous dependence is sensitive to the external strain and it can be tuned easily by the external compressive strain.

†Corresponding author E-mail: hongjw@bit.edu.cn (J Hong); Tel: +86 010 68915917

**Introduction**

With the successful exfoliation of graphene in the year of 2004[1], the research society has witnessed the rapid development of the two-dimensional (2D) materials with emerging fascinating physical properties[2,3]. The size-effect induced quantum phenomenon, such as Dirac cone, long-range magnetic ordering, topologically protected band structure, *etc.* have boosted the novel materials research tremendously, including experimental synthesis and theoretical designing[4-6]. Combining the 2D property with tradition ferroic orders, such as magnetic order or ferroelectricity, will improve the potential applications of novel 2D materials. Especially the controlled manipulation of polarization in ferroelectric (FE) materials in the presence of external fields are critical for the potential applications in non-volatile memories, actuators, and sensors[7-9]. Driven by technological demand for device miniaturization, exploration of the ferroelectric properties at two-dimensional (2D) limit has become more urgent[8,10-13]. The spontaneous electric polarization, Curie temperature ($T_c$) and atom distortions are usually thickness-dependent, and the electric polarization decreases monotonically as the film thickness decreasing, then disappeared at a critical value, due to the existence of the depolarizing electrostatic field and surface energy caused by surface reconstruction[13-15].

Recently, Chang *et. al.* discovered that the atomically thin SnTe film exhibits robust in-plane ferroelectricity, which is different from the surface structures sliced directly from the bulk phase[16]. Compared to the low $T_c$=98 K of bulk SnTe, the $T_c$ value of monolayer SnTe was greatly enhanced to 270 K. In addition, for the thicker films including 4- to

8-atomic layers, the $T_c$ is even higher than the room temperature[16]. However, the origin of such phenomenon, and its thickness dependent of geometric structure, barrier height and $T_c$ are yet thoroughly investigated. In this work, we systematically investigated the thickness dependent ferroelectric properties of 2D SnTe, which stabilized in a special periodic bilayer structure. Based on the double-well potential analysis, a non-monotonous dependence of $T_c$ on thickness has been revealed, and surprisingly, a maximum polarization can be achieved at 6 atomic-layers thickness. In addition, the barrier height of the double-well potential has been studied in the presence of an external biaxial strain, which shines some lights on the design of the future nanoscale ferroelectric devices.

**Computational method**

All computations were carried out in the frame work of density functional theory (DFT) as implemented in the Vienna Ab initio simulation package (VASP)[17]. The local-density approximation (LDA) is used for the exchange-correlation. The projector augmented wave method was employed to model the ionic potentials[18]. A vacuum space of 10 Å was introduced to avoid interactions between slabs. All the atomic positions and lattice parameters are allowed to relax until the calculated forces less than 0.001 eV/Å, while the electronic minimization was applied with a tolerance of $10^{-6}$ eV. The phonon dispersions were calculated with VASP and Phonopy[19], using 4×4×1 supercells. Based on the convergence test, we used a plane-wave cutoff of 400 eV for all calculations. The Monkhorst-pack k-point sampling[20] was used for the Brillouin zone integration: 16×16×1 for the unit cell and 4×4×1 for the supercell, respectively.

**Results and discussions**

We chose the 2D SnTe slabs by cutting the bulk SnTe along (001) plane, then fully relaxed the slabs, as presented in Fig. 1a. In order to check the stability of 2D SnTe slabs, the formation energy was calculated, which is given by the cohesive energy difference $E_f=[E_{slab}(N)-NE_{bulk}]/N$, where $E_{slab}(N)$ and $E_{bulk}$ are the total energies of the 2D SnTe and bulk SnTe, $N$ is the number of atomic layers, as shown in Fig. 1(b). It can be seen that the formation energy shows oscillations with increasing the atomic layers. Interestingly, the formation energy increases for the SnTe slab with odd-number-layer, then decreases for the SnTe slab with even-number-layer, which indicates that the SnTe slabs prefer a periodical bilayer structure. The energy difference between the odd-number-layer and even-number-layer gradually reduces. This result is consistent with the experimental observation, where only structure with even-number-layer structures are preferable and have been successfully synthesized[16]. Therefore, we only consider the SnTe slab with even-number-layers in our work.

Figure 1(c) shows the interlayer spacing for the even-number-layers structure, from which the interlayer spacing $d_i$ (i=2, 4 ,6…) of the 2D SnTe is larger than that of the bulk SnTe, while the interlayer spacing $d_j$ (j=1, 2 ,3…) is smaller than that of the bulk SnTe. Thus, the SnTe slabs automatically transformed into periodical bilayer structure with a few atomic layers. Besides, the layer spacing ($d_i$ and $d_j$) is closer to that of the bulk with the increase of the thickness. Interestingly, it can be seen that the outermost layer spacing ($d_1$) only has slight decrease for the slab with 12 atomic layers, indicating the surface effect is still significant in such slab. In addition, the changes of in-plane lattice constants as increasing the thickness are presented in Table 1, which indicates the thinner 2D SnTe is, the larger lattice constants are. It suggests that the surface effect leads to a lattice expansion.

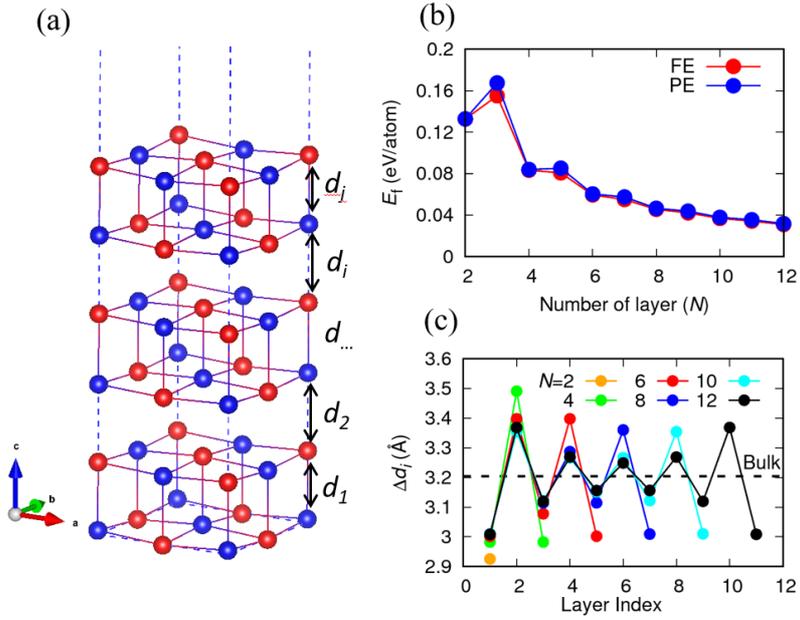

Fig. 1. (a) Geometric structure of 2D SnTe. (b) Formation energies for different number of atomic layers. (c) Interlayer spacing for different number of atomic layers, the dashed line indicates the atomic layer distance of bulk.

Traditionally, the tetragonality ratio is used as an indicator to reveal the ferroelectric properties. For example, the tetragonality ratio and ferroelectric polarization decreases simultaneously with the decreasing the film thickness[21,22]. The in-plane tetrangonality ratio of 2D SnTe with differernt thickness is summarized in Table 1. It is clear that the in-plane tetragonality ratio of the ferroelectric phase has an unusual nonmonotonous thickness dependence behavior, with the maximum tetragonality ratio occurss at 8-atomic-layer structure. The nonmonotonous thickness dependence character indicates an unusual ferroelectric property of 2D SnTe.

Table 1. In-plane lattice constants of ferroelectric structure and the tetragonality ratio *a/b* for different number of atomic layers. The polarization is along *a* direction ([110] direction)

| Lattice constant (Å) | Number of Atomic Layer | | | | | |
|---|---|---|---|---|---|---|
| | 2 | 4 | 6 | 8 | 10 | 12 |
| *a* | 4.574 | 4.572 | 4.570 | 4.569 | 4.566 | 4.565 |
| *b* | 4.556 | 4.536 | 4.529 | 4.527 | 4.526 | 4.526 |
| *a/b* | 1.0040 | 1.0079 | 1.0090 | 1.0094 | 1.0088 | 1.0086 |

We calculated the phonons dispersions of SnTe paraelectric slabs ($N=2$ and 3), as shown in Fig.2. It can be seen that there are soft phonon modes at $\Gamma$ point for both slabs. However, for the slab with odd atomic layers ($N=3$), there are additional soft modes at M point, indicating the stronger instability for the slabs with odd atomic layers. Next we will focus on the soft polar modes at $\Gamma$ point which will be frozen and drives paraelectric slab to ferroelectric state when temperature is lower than the the Curie temperature $T_c$. The atomic vibration pattern of soft polar mode at $\Gamma$ is shown in Fig.2b, in which Sn atoms and Se atoms vibrate oppositely along [100] direction. (There is another degenerated mode along [010], not shown here).

The frozen-phonon potential of the soft mode at $\Gamma$ following the vibration modes along [100] (Fig.2b) and [110] (Fig.2d) were calculated, as shown in Fig.3. Both modes exhibit a "double well" behavior (the potentials are symmetric and only the positive side is shown). We find that the [100]-mode has a shallower barrier energy (labeled as $E_H$-[100]), while the [110]-mode has a deeper one (labeled as $E_H$-[110]). It reveals that the ferroelectric structure with [110]-mode frozen from the paraelectric structure is the ground state structure and the ferroelectric polarization is along [110] direction, which agrees with the experimental observation[16].

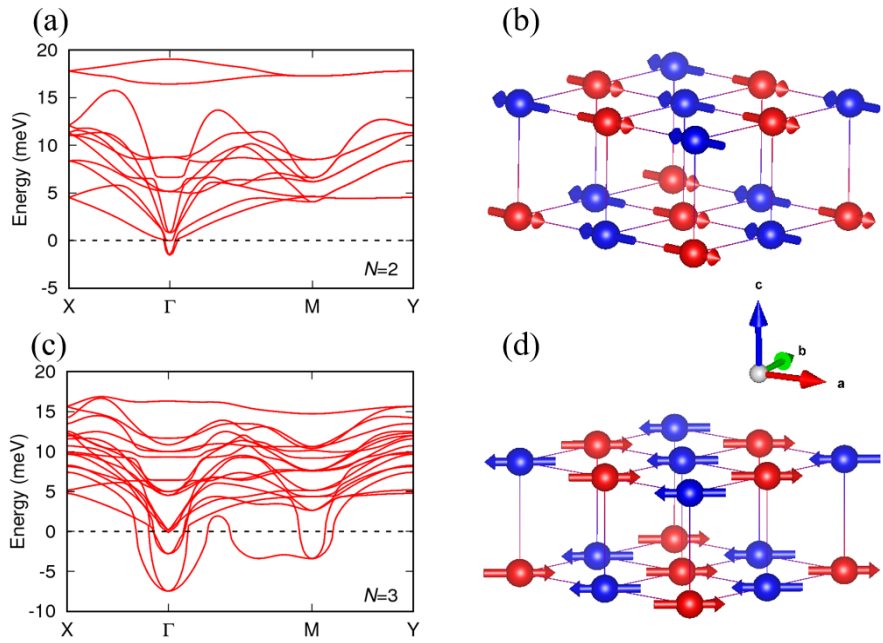

Fig. 2. Phonon dispersion for monolayer 2D SnTe. (a) Atomic layer $N=2$. (b) Atomic layer $N=3$. (c), (d) Vibration modes at Gamma point along [100] and [110] direction for monolayer 2D SnTe.

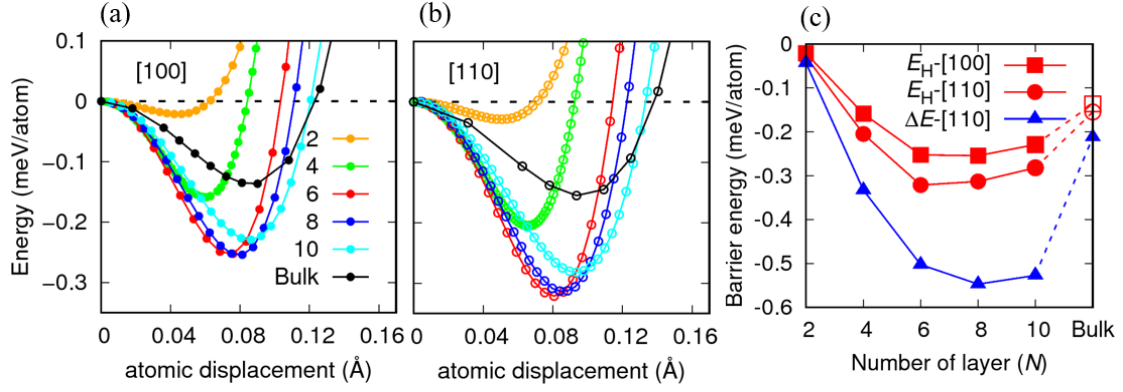

Fig. 3. Frozen-phonon potential for the vibration modes along [100] (a) and [110] (b) directions for different number of atomic layers. (c) The barrier energy ($E_H$) for different number of atomic layers. $E_H$ is extracted from (a) and (b) while $\Delta E$-[110] is the energy difference between the fully relaxed 2D SnTe slabs with initio distortion along [110] direction and the paraelectric phase.

The barrier energy was extracted from the double well potential, as shown in Fig.3c. It can be clearly seen that the barrier energy initially increases as thickness increases, reaching a maximum for 6 atomic-layers structure, and then decreases to the bulk value as thickness increases further. It displays an unusual nonmonotonous thickness dependence character which is similar to the thickness dependence of the tetragonality ratio of the 2D SnTe (Table 1). However, there is a slight difference: the minimum barrier energy is 6-atomic-layers while the maximum tetragonality ratio is 8-atomic-layers. This difference is probably due to the barrier energy extracted from the double-well potential did not take into account the volume change during the paraelectric-ferroelectric phase transition. To further prove it, we fully relaxed the ferroelectric phase structure of 2D SnTe slabs and obtained the energy difference $\Delta E$-

[110] between fully relaxed ferroelectric and paraelectric structures. We find that nonmonotonous thickness dependence character is barely influenced, except that the barrier heights are all enhanced. Interestingly, the minimum barrier energy is changed from the 6-atomic-layer to 8-atomic-layer, which is coincided with the maximum of the tetragonality ratio.

The atomic distortion in each atomic layer is shown in Fig. 4(a). It can be found that when $N=2$, the atom distortion is equal with each other, and both of them are smaller than that of the bulk SnTe atom distortion. However, when $N>2$, the atom distortion in each atomic layer are no longer equal to each other. The odd layers from surface (1st,3rd, …) has larger distortion than that of bulk while the even layers (2nd ,4th, …) has smaller distortion, which indicates that the interlayer interaction has great impact on the atom distortions. Moreover, we find that the atom distortions of the surfaces of 2D SnTe are much larger than the atom distortions of bulk SnTe (except for $N=2$). With the increase of the thicknesses, the atom distortions of the 2D SnTe gradually decrease. When the thickness of 2D SnTe is large enough, the atom distortions of 2D SnTe will be equal to the atom distortion of bulk SnTe.

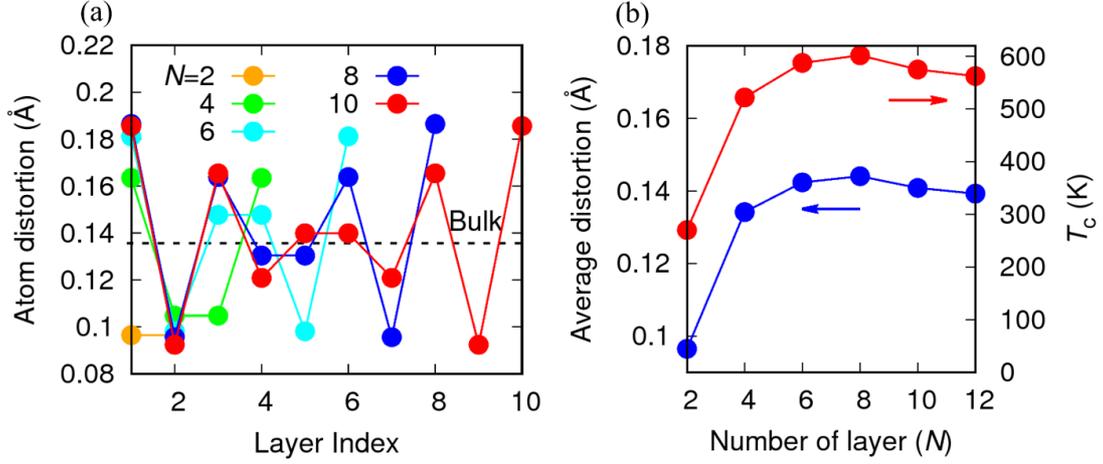

Fig. 4. (a) Atom displacement in each atomic layer for different number of atomic layer 2D SnTe. (b) Average atomic displacement and estimated Curie temperature for different number of atomic layer 2D SnTe.

We further calculated the average atomic distortions to investigate the thickness-dependent Curie temperature ($T_c$), as shown in Fig. 4(b). It can be seen that the average atomic distortion also shows a nonmonotonous thickness dependence, with the maximum distortion for 8-atomic-layer slab, consistent with the previous thickness dependence of tetragonality ratio and energy barrier. We then estimated the $T_c$ based on the average atom distortion with the following empirical formula: [23]

$$kT_c = (1/2)\, \mathrm{K}\, (\Delta z)^2$$

where k is the Boltzmann constant, $\mathrm{K}$ has the dimensions of a force constant, $\Delta z$ is the atom distortion. We chose the experimental Curie temperature ($T_c$ = 270 K [16]) of SnTe slab with $N$=2 to fit the $\mathrm{K}$ and then estimated the $T_c$ for the slabs with different thickness, as shown in Fig.4b. It can be seen that the SnTe film with 8 atomic layers possesses the highest Curie temperature ($T_c$ =602K). It should be noted that this is the preliminary estimation of Curie temperature based from the experimental $T_c$ from the thinnest SnTe film (*N*=2). Some other theoretical methods were also introduced to investigate the thickness dependence of the Curie temperature of SnTe and similar trend was obtained recently[24,25].

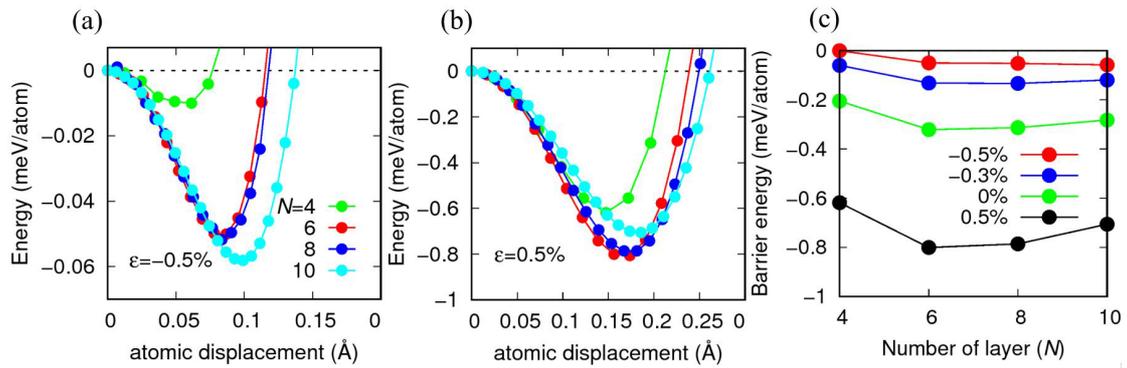

Fig. 5. Frozen-phonon potential with atomic vibration along [110] direction under the biaxial strain -0.5% (a) and 0.5% (b). (c) Barrier energy of SnTe slabs in the presence of different external strain.

Since strain has great impact on the ferroelectric properties[26,27], we investigated the effect of the biaxial strain on the barrier energy by using the frozen-phonon method to calculate the double-well potential, as illustrated in Fig. 5. We find that the tensile strain can greatly lower the barrier energy. For example, when 0.5% tensile strain are applied on the 2D SnTe, the barrier energy is lowered by about 3 times, indicating the in-plane ferroelectricity is enhanced by tensile strain. Moreover, the tensile strain has no impact on the nonmonotonous thickness dependence character of the barrier energy. On the contrary, the compressive strain enlarges the barrier energy. For example, when -0.5% compressive strain is applied on the 2D SnTe, the barrier energy decreases and the double well potential is very shallow, indicating the in-plane ferroelectricity is suppressed significantly under compressive strain. Interestingly, the compressive strain is found to change the nonmonotonous thickness dependence of the barrier energy. From Fig. 5(c), we can see that the 6 atomic-layer structure has the lowest barrier energy without external strain, but it changes to 8-atomic layer with -0.3% compressive strain and 10 atomic-layer structure with -0.5% compressive strain.

**Conclusion**

In conclusion, by using DFT calculations, we find 2D SnTe automatically transform into periodically bilayer structures, due to the surface effect. The barrier energy of the double-well potential first reduces and then increases, as the number of atomic layers increases, showing a nonmonotonous thickness dependence. Based on the atomic distortion analysis, the Curie temperature of the 2D SnTe is found to be higher than its bulk counterpart, and the Curie temperature on the thickness dependence exhibits an unusual nonmonotonous behavior as well, featuring a pronounced maximum at 8 atomic-layers. Moreover, the biaxial strain has significant impact on the barrier energy of the double-well potential for the 2D SnTe and Curie temperature. Our work helps to understand the thickness dependent ferroelectric properties of 2D SnTe, and shine lights on the potential application for next-generation nanoelectronic devices under external strain.